\PassOptionsToPackage{square,comma,numbers,sort&compress}{natbib}
\documentclass[12pt]{article}
\usepackage[T1]{fontenc}
\usepackage{natbib}
\usepackage[utf8]{inputenc}
\usepackage{lmodern}
\usepackage{amsmath}
\usepackage{amssymb}
\usepackage{graphicx}
\usepackage{subcaption}
\usepackage{color}

\title{Cosmic censorship conjecture in a general Kerr-Newman black hole}
\author{H. Khodabakhshi$^{1,2}$ , F. Shojai$^{1,3}$  \\
\small$^1$Department of Physics, University of Tehran\\
\small$^2$School of Particles and Accelerators, \\ \small Institute for Research in Fundamental Sciences (IPM),\\
\small P.O.Box 19395-5531 Tehran, Iran\\
\small$^3$Foundations of Physics Group, School of Physics, \\ \small Institute for Research in Fundamental Sciences (IPM)\\
 }
\date{}
\begin{document}
\maketitle
\begin{abstract}
Using some probes, the violation of cosmic censorship conjecture in a general Kerr-Newman black hole is investigated. The result depends on many factors, like the relative sign of charge and rotation direction of the probe and black hole. Moreover the comparison of the angular momentum of the black hole and its charge has an impressive effect. Considering all these together, we have found the range of the angular momentum, energy and charge of the probe for which the event horizon disappears. We have found that cosmic censorship conjecture violation is possible only for near extremal black hole if the parameters of the probe are too large and fine tuned. Taking into account the hoop conjecture, we see that the cosmic censorship conjecture is respected for a general kerr-Newman black hole subject to a large enough number of falling particles and also field quanta.
\end{abstract}

\section{Introduction}

In general relativity (For recent overviews, see, e.g., [1-3] and references therein) according to the weak cosmic censorship conjecture (CCC) [4,5] for a geodesically complete and asymptotically flat space–time, the evolution of matter fields satisfying the null energy condition cannot lead to a naked singularity. This means that any black hole (BH) singularities formed by gravitational collapse of matter remain hidden behind an event horizon [6-8]. For a general Kerr–Newman (KN) BH, the CCC is usually tested by adding charged and rotating particles or some fields and seeing if the BH horizon is destroyed or not. Also, there are potential probes for studying the physics close to a supermassive BH. For example the orbit of S stars around the BH Sgr (Sagittarius) A*, and attempts to use them as tests for GR can be found in [9-19]. It is worth mentioning that, recently, an interesting line of research emerged pointing out that Earth-like forms can develop even in rogue planets formed in-situ and orbiting supermassive BHs thanks to the irradiation of the accretion disk under certain circumstances. Orbits of such planets play an important role in such a context [20-25].

 It has been claimed that CCC can be violated for an extremal KN BH with fine tuning of the particle's energy \cite{16}, and  further discussion along these lines takes into account a cosmological constant 
  \cite{17,18,19}.  Wald has shown \cite{12} that by considering terms linear in the angular momentum, energy and charge of a particle falling toward an extremal KN BH, the horizon cannot be destroyed. Jacobson and Hubney \cite{14,15} {suggested} that the CCC could be violated for a slightly non extremal BH if every physical quantity is expanded in powers of small parameters up to the appropriate order. However according to Wald \cite{13}, taking into account all effects on such black holes that are second-order in the charge and angular momentum, no over-charging or over-spinning can occur in Gedanken experiments of the Hubeny type.
  
Investigations of CCC {violation} using a scalar field {probe} \cite{Bek} has also been of recent interest \cite{a1,a2,a3,a4}. Neglecting back-reaction effects,  lower and upper bounds for mode 
frequency were obtained, assuming that the scalar field is absorbed by the BH and the event horizon is destroyed by it \cite{a4}. Using these frequencies, it was shown that  CCC violation is approximately possible for an extremal KN BH. 

In this paper, we  investigate CCC violation in a general KN BH using both particle and field probes. It is convenient to define some alternative dimensionless parameters for the BH, for the particle and   for the scalar field. We shall do this in the next section and write the necessary conditions for violating the CCC. In section 3 we confine our attention to three simple special cases -- Kerr, RN and extremal KN BHs -- and investigate  whether or not it is possible that the event horizon  can disappear  for valid ranges of particle/field parameters.  We then extend our discussion to a general KN BH.   In section 4 we provide a   more precise analysis, discussing whether or not  CCC can be violated taking into account the hoop conjecture for a general KN BH or not.  At the end in section 5, we present a summary of the main conclusions.

\section{Destroying the event horizon of a general KN BH}

In general relativity, the asymptotically flat stationary BH solution of the Einstein-Maxwell equations is described by the KN metric 
\begin{align}\label{KN*}
dS^2=-\frac{\rho^2\Delta}{\Sigma^2}dt^2+\frac{\Sigma^2 \sin^2{\theta}}{\rho^2}(d\phi-\bar{\omega}dt)^2+\frac{\rho^2}{\Delta}dr^2+\rho^2d\theta^2 
\end{align}
and the following vector potential of the electromagnetic field 
\begin{align}
A_t=&\frac{-Qr}{\rho^2} && A_{\phi}=\frac{Qar\sin^2{\theta}}{\rho^2} & A_r=A_\theta=0
\label{As}
\end{align}
where $\rho^2=r^2+a^2\cos^2{\theta}$, $\Delta=r^2+a^2+Q^2-2Mr$, $\Sigma^2=(r^2+a^2)^2-a^2\Delta\sin^2{\theta}$ and $\bar{\omega}=2Mra/\Sigma^2$. The parameters $M$, $Q$ and $a=J/M$ correspond to the mass, charge and the angular momentum (per mass) of the BH. 

The outer horizon is given by the greater root of $\Delta=0$, which is $ r_+=M+\sqrt{M^2-Q^2-a^2} $. The thermodynamic properties of the BH depend on this parameter, with
\[
  \Omega=\frac{a}{r_+^2+a^2}  \hspace{1cm} {T}=\frac{r_+-M}{2\pi(r_+^2+a^2)}
\]
\begin{equation}\label{fori}
\Phi=\frac{r_+Q}{r_+^2+a^2}    \hspace{1cm}   \mathit{S}=\pi(r_+^2+a^2)
\end{equation}
being the angular velocity,   temperature, electric potential, and Bekenstein-Hawking entropy of the BH respectively.  The first law reads \cite{a5}
\begin{equation}
  \delta M=T\delta \mathit{S}+\Omega\delta J+\Phi\delta Q
\end{equation}
where $\delta$ denotes variations of the KN BH's parameters.  
Assuming the weak energy condition, the second law of thermodynamics holds
\begin{equation}\label{x2}
\delta\mathit{S}\geq 0
\end{equation}
or in other words the BH entropy cannot  decrease.

 Varying $r_+$ and $a$ and substituting the result into (\ref{x2}), one gets
\begin{equation}
\delta M\geqslant \frac{a\delta J+Qr_+ \delta Q}{a^2+r_+^2}
\label{1*}
\end{equation}
Hence if the black hole is slightly perturbed by a probe, the change of BH's parameters should satisfy (\ref{1*}). Now consider a particle  with energy $E$, charge $q$, and angular momentum $L$ falling into a KN BH. If the particle passes through the outer horizon, then the mass, charge and the angular momentum of the BH change. Meanwhile some energy of the particle  is lost by radiation. Thus according to the conservation laws
\begin{align}
  \delta M<E &,&\delta J=L&,&\delta Q=q
  \label{x1}
\end{align}
Here we have assumed that the initial spin of the particle is along the axis of symmetry of the BH. By symmetry, it can be easily seen that it remains parallel to the axis of the BH. Thus after absorbing the particle,  the BH is still axisymmetric and any radiation from the particle does not carry away any angular momentum \cite{12}. Comparing the above inequality with (\ref{1*}), gives the following lower bound for the particle's energy\footnote{Equivalently, one can say that for an infalling particle, the conserved energy and angular momentum are 
\[
E=-(m u_\mu+q A_\mu)(\frac{\partial}{\partial t})^\mu \hspace{1cm} L=(mu_\mu+q A_\mu)(\frac{\partial}{\partial \phi})^\mu
\]
where $ u^{\mu} $ is its four-velocity. Given $E$ and $L$, we may eliminate $\dot{ \phi}$ in the above relations and obtain $\dot{t}$ as 
\[
\dot{t}=\frac{g_{\phi\phi}(E+q A_t)+g_{t\phi}(L-q A_{\phi})}{m(g_{t\phi}^2-g_{\phi\phi}g_{tt})}
\]
To get a future pointed velocity, one should have $\dot{t}>0$ which means that the numerator of the above relation is positive. This gives a lower bound for $E$ where can be evaluated at the outer horizon of  the BH, yielding (\ref{1^}).}
\begin{equation}
E\geqslant \frac{aL+qQr_+}{a^2+r_+^2}.
\label{1^}
\end{equation}
The right side of (\ref{1^}) defines a potential barrier. If the particle's energy does not satisfy (\ref{1^}), the above potential barrier prevents it from falling into the BH. 

 Furthermore, before  the particle enters the BH, the  horizon existence condition 
\begin{equation}
M^2 \geq Q^2+a^2
\label{2*}
\end{equation}
must be  satisfied, where equality holds for an extremal KN BH. Since we are interested in destroying the horizon by throwing a particle into it, we must have\footnote{It should be noted here that expanding relation (\ref{3*}) up to the first order in $\delta M/M$, $\delta Q/Q$ and $\delta J/{aM}$ gives
\begin{equation}\label{ty}
M^2-a^2-Q^2<Q\delta Q+\frac{a}{M}\delta J-\frac{M^2+a^2}{M}\delta M
\end{equation}
Since the right-hand side is a differential expression, the BH would be extremal or near extremal if \textbf{its} parameters satisfy the above condition. Moreover according to (\ref{2*}), the left hand side of (\ref{ty}) is non-negative. Thus
\begin{equation}
\delta M-\frac{a \delta J}{M^2+a^2}-\frac{QM}{M^2+a^2}\delta Q\leq0
\end{equation} 
This is  a violation of the second law of thermodynamics, \textbf{eq. (\ref{x2})}. This means that up to  first order  in the small parameters introduced above, the CCC can not be violated for an extremal BH \cite{13}.} 
\begin{equation}
(\delta M+M)^2-(Q+q)^2-\big(\frac{aM+L}{M+\delta M}\big)^2<0
\label{3*}
\end{equation}

However we instead shall assume that
\begin{equation}\label{do}
(E+M)^2-(Q+q)^2-\big(\frac{aM+L}{M+E}\big)^2<0
\end{equation}
since (\ref{do}), together with the first inequality in (\ref{x1}), results in (\ref{3*}). On the other hand, (\ref{do}) is equivalent to 
\begin{equation}\label{simple}
E <\sqrt{\frac{(Q+q)^2}{2}+\sqrt{
(aM+L)^2+\frac{(q+Q)^4}{4}}} - M
\end{equation}
Therefore the energy of infalling particle should be smaller than the right hand side of (\ref{simple}) in order for the horizon to disappear.
The inequalities (\ref{1^}) and (\ref{simple}) subject to the constraint (\ref{2*}),
give the allowed range for the energy, charge and angular momentum of the particle to violate   CCC. For an extremal KN BH, comparing the upper and lower bounds of energy,  (\ref{1^}) and (\ref{simple}), and expanding the result up to the second order in $q$ and $L$, we obtain the allowed ranges of the particle's energy $E$ \cite{16}. This can be done even analytically for some special choices, the extremal Kerr or extremal RN BH \cite{12,14,15,16}. 

However, here, we focus on a general (not necessarily extremal) BH without any expansion in small parameters.  To do this, it is more convenient to employ  the dimensionless parameters 
\begin{equation}
 \epsilon=1+\frac{E}{M} \qquad \eta=1+\frac{L}{aM} \qquad \xi=1+\frac{q}{Q} \qquad \lambda=\frac{Q^2}{M^2}
 \qquad  \gamma=\frac{a^2}{M^2}
\end{equation}
associated with the particle, where the latter two parameters measure to what extent the BH is RN-like and Kerr-like respectively.  Using these, we write
the inequalities  (\ref{1^}) and  (\ref{simple}) as
\begin{align}
\gamma\eta+\lambda\epsilon-(\lambda+\gamma)+(1+\sqrt{1-(\gamma+\lambda)})(\lambda\xi-2\epsilon-\lambda+2) \leqslant  0
\label{4*}
\end{align}
\begin{equation}
\epsilon^2<\frac{\lambda\xi^2}{2}+\sqrt{\frac{\lambda^2\xi^4}{4}+\gamma\eta^2}
\label{6*}
\end{equation}
Also the constraint (\ref{2*}) can be written as 
\begin{equation}
\gamma+\lambda\leq1
\label{7*}
\end{equation}
where equality holds for the extremal case. Clearly the charge of the particle and the BH can be of the same or opposite signs and also they can rotate in the same or opposite directions. Moreover, we must have $|\xi|$, $|\eta|$ and $\epsilon \sim 1$ to \ justify the particle assumption. In the next section, by specifying the input parameters $\gamma$ and $\lambda$ such that the constraint (\ref{7*}) is satisfied, we plot (\ref{4*}) and  (\ref{6*}) in three dimensional space of $\xi$, $\eta$ and $\epsilon$ to find for which intervals of these  parameters is   CCC  violated.


We next follow \cite{Bek} and consider a charged scalar field  propagating  in the KN BH space-time. A given scalar mode has frequency $\omega$ and azimuthal quantum number $m=0,\pm 1,\pm 2,...$. Moreover assume that its dependence on $r$ and $\theta$ (in Boyer-Lindquist coordinates) is such that it behaves like an ingoing wave near the BH  horizon, whereas it has both ingoing and outgoing components far from it. A simple calculation of the stress-energy tensor of this field shows that
\begin{equation}\label{lf}
\frac{L^{(f)}}{E^{(f)}}=\frac{m}{\omega}
\end{equation}
and
\begin{equation}\label{qf}
\frac{q^{(f)}}{E^{(f)}}=\frac{\mu}{\omega}
\end{equation}
where $E^{(f)}$ and $L^{(f)}$  are the respective  energy and angular momentum of the scalar field and $\mu=q^{(f)}$ is a classical charge parameter.  We assume that the scalar field is absorbed and also scattered by the BH and moreover the total angular momentum and charge of the scattered part are zero. This yields
\begin{equation}\label{ff}
\delta M< E \hspace{2cm} \frac{\delta Q}{E}=\frac{\mu}{\omega} \hspace{2cm} \frac{\delta J}{E}=\frac{m}{\omega}
\end{equation}
which is identical to (\ref{x1}) obtained for a particle probe. If the second law of BH thermodynamics holds, then clearly (\ref{1*}) is
again satisfied. Moreover using (\ref{ff}) and the conservation of charge and angular momentum,  we quickly find that
\begin{equation}\label{mai}
E\geqslant \frac{a L^{(f)}+Qr_{+}q^{(f)}}{a^2+r_+^2}
\end{equation}
or, using (\ref{lf}) and (\ref{qf}) in (\ref{mai}) 
\begin{equation}\label{re}
\omega\geq \frac{am+\mu Qr_+}{a^2+{r_+}^2}
\end{equation}
This shows that the mode is attenuated  due to interaction of the field with the BH\footnote{Mode amplification occurs when   $\omega <m\Omega+\mu\Phi$. This together with the second law of thermodynamics, means that $\delta M<0$, i.e. energy is extracted from the BH.}.
According to (\ref{re}), there is a minimum mode frequency, $\omega_{min}$ above which the BH absorbs the scalar field. Otherwise it would be scattered to infinity. Following what was done for the particle probe,  there is also another inequality that gives an upper bound for mode frequency, $\omega_{max}$ and comes from assuming the interacted BH has been overspun/overcharged to a naked singularity\footnote{In \cite {a4}, for a neutral scalar field, both sides of the inequality $\omega_{min}<\omega_{max}$ are expanded up to the second order in powers of $\frac{\omega}{M} \ll 1$. This gives the condition $a^2<\frac{1}{3}$ to overspin an extremal BH.}
\begin{equation}\label{wsimpl}
\omega<\sqrt{\frac{(Q+\mu)^2}{2}+\sqrt{(aM+m)^2+\frac{(\mu+Q)^4}{4}}}-M
\end{equation}

 Putting the above results together, we see that the relations (22) and (23) for the scalar field are respectively similar to (8) and (14) for the particle. Hence (15) and (16), with appropriate parameter redefinitions, apply to the scalar field as well. However, it should be noted that the mode number and field charge change discontinuously.
 
In the next section, first we confine our attention to some special BHs and then discuss a general KN BH.
\section{Bounds on the probe's parameters}
A particularly interesting special case is the Kerr BH for which $ \lambda=0 $ and there is no need to introduce $\xi$. So we can write (\ref{4*}), (\ref{6*}) and (\ref{7*}) as 
\[\gamma(\eta-1)+2(1+\sqrt{1-\gamma})(1-\epsilon)\leqslant  0\]
\begin{equation}
\epsilon^2<\sqrt{\gamma}\eta \hspace{1.5cm}\gamma<1
\label{ineq3}
\end{equation}
In this case, the three dimensional parameter space consists of $\gamma$ which is related to the BH and $\epsilon$ and $\eta$, the probe's parameters.  As mentioned before, the probe and the BH could have the same (or opposite) direction of rotation, so $\xi$ and $\eta$ could be greater (or smaller) than unity. Plotting the above inequalities, we have easily found that there is no solution over the valid ranges of parameters, $0<\gamma<1$, $0.9<\eta<1.1$ and $1<\epsilon<1.1$. Thus the CCC is not violated for a  Kerr BH.

For an  extremal Kerr BH, $ \gamma=1 $ and thus (\ref{ineq3}) gives 
\begin{equation}
\eta-2\epsilon+1\leqslant  0 \hspace{1.5cm} \epsilon^2<\eta
\label{ineq5}
\end{equation}
These inequalities have no solution, which means that the CCC  is  not violated for an extremal Kerr BH \cite{14,15,16}.

Another special case is the RN BH for which $ \gamma=0 $ and there is no need to introduce $\eta$. Thus the inequalities (\ref{4*}), (\ref{6*}) and (\ref{7*}) become
\[\lambda(\epsilon-1)+(1+\sqrt{1-\lambda})(\lambda\xi-2\epsilon-\lambda+2)\leqslant  0\]
\begin{equation}
\epsilon^2<\lambda\xi^2 \hspace{1.5cm} \lambda<1
\label{3**}
\end{equation}
Once again the space of parameters is three dimensional. $\lambda$ characterizes the BH and $\epsilon$, $\xi$ are the probe's parameters. Plotting the above inequalities, we again see that there is no solution in the allowed  parameter ranges. Therefore the CCC is always valid for a RN BH. 

For the extremal RN case $ \lambda=1 $, two above inequalities can be written as
\begin{equation}
\xi-\epsilon\leqslant 0 \hspace{1.5cm}\epsilon^2<\xi^2
\label{xx1}
\end{equation}
It is clear that  the above inequalities are inconsistent. Thus the CCC is not violated for the extremal RN BH \cite{14,15,16}.\\
\begin{figure}
	\begin{center}
		\includegraphics[height=12cm]{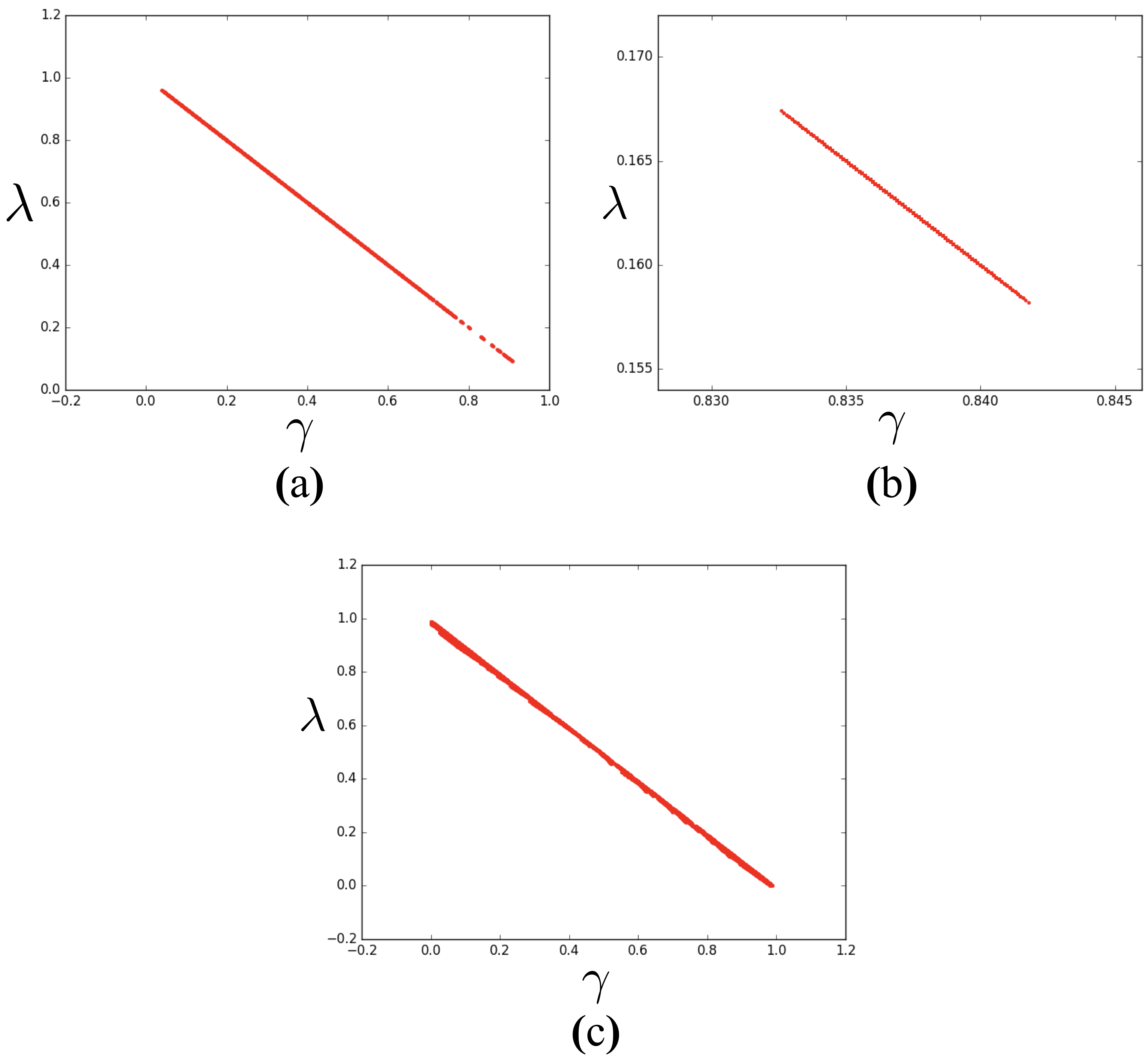}
		\caption{\small{CCC is violated in the specified $\gamma$ and $\lambda$ in (a) and (b) for an extremal KN BH and in (c) for non extremal case. In  (a) and (c): $1<\epsilon<1.1$, $0.9<\xi<1.1$ and $0.9<\eta<1.1$ and in  (b): $1.00128<\epsilon<1.0013$, $1.0095<\xi<1.0097$, $1.00095<\eta<1.00097$. (c) shows that CCC is not violated for $\gamma+\lambda$ sufficiently less than one.}}
		\label{fig:0}
	\end{center}
\end{figure}

For an extremal KN BH, by substituting $ \lambda=1-\gamma $ into (\ref{4*}) and (\ref{6*}), we obtain 
\begin{equation}
\eta+\frac{1-\gamma}{\gamma}\xi-\frac{1+\gamma}{\gamma}\epsilon+1\leqslant  0
\label{12*}
\end{equation}
\begin{equation}
\epsilon^2-\frac{(1-\gamma)\xi^2}{2}-\sqrt{\frac{(1-\gamma)^2\xi^4}{4}+\eta^2\gamma}<0
\label{13*}
\end{equation}
 In Fig. 1a, the CCC violating range of parameters $\gamma$ and $lambda$ is shown, when the particle parameters areintheirvalidrange, $1<\epsilon<1.1$, $0.9<\zeta<1.1$ and $0.9<\eta<1.1$.

Also, in order to compare charge and angular momentum of BH, we take $\gamma$ as an input parameter and then consider three cases: $ \gamma>\lambda $, $ \gamma\sim \lambda $ and $ \gamma<\lambda $. For each case, again, $\xi$ and $\eta$ could be a little greater or smaller than unity. 
\begin{figure}
	\begin{center}
		\includegraphics[height=15cm]{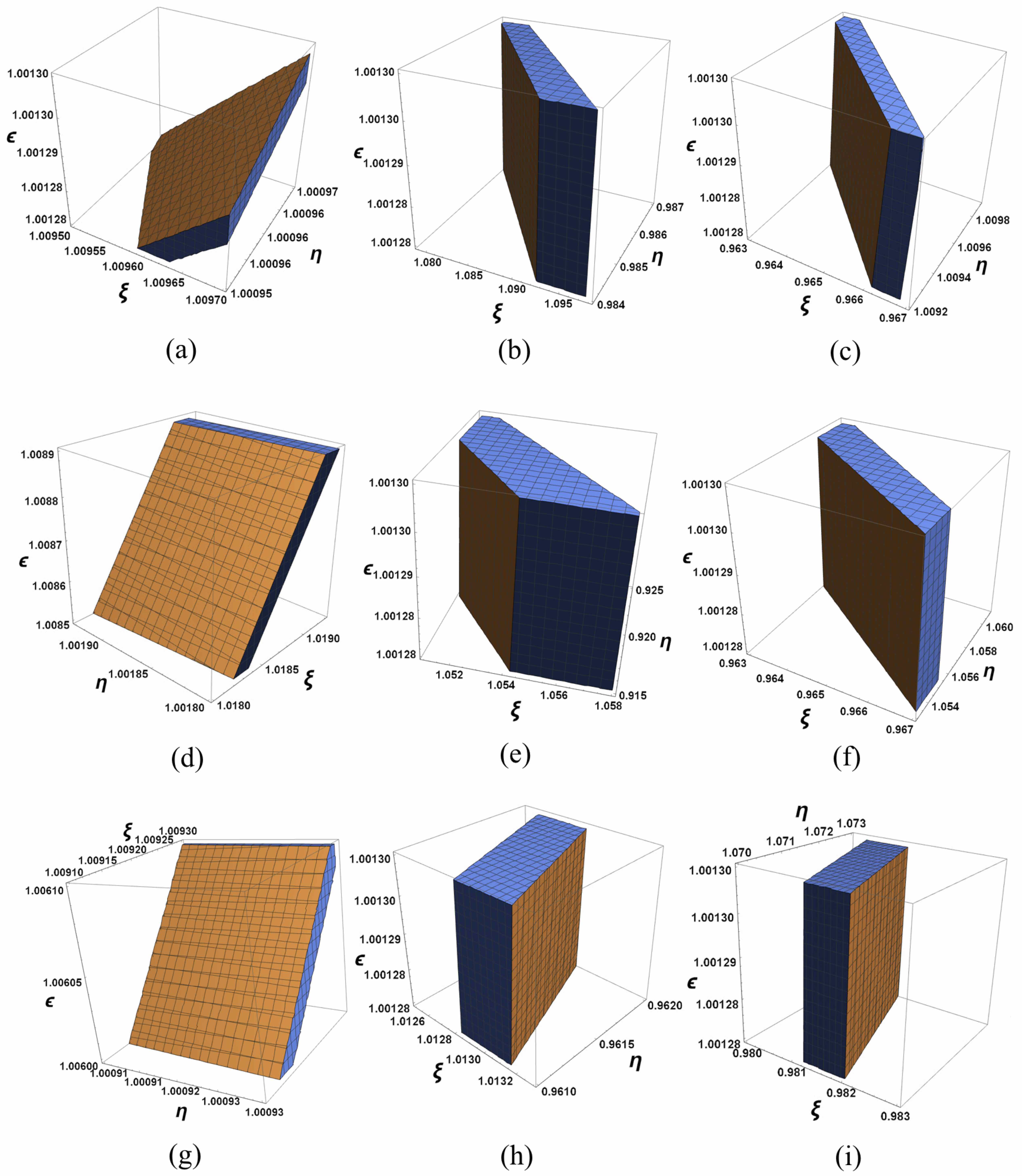}
		\caption{\small{CCC is violated in the specified region of (a)-(i) for an extremal KN BH. In diagrams (a), (b) and (c), $\gamma>\lambda$ and $\gamma=0.8383$, in diagrams (d), (e) and (f), $\gamma\sim\lambda$ and $\gamma=0.3939$ and in diagrams (g), (h) and (i), $\gamma<\lambda$ and $\gamma= 0.2222$. }}
		\label{fig:1}
	\end{center}
\end{figure}
Contrary to the previous cases, we find  allowed   parameter ranges  for violating  CCC. The result is plotted in figure~\ref{fig:1} for several special cases of $\gamma$.

consider for example is~\ref{fig:1}a,  where parameter values within the ranges
\begin{equation}\label{ab}
1.00128<\epsilon<1.0013\hspace{.55cm}1.0095<\xi<1.0097\hspace{.55cm}1.00095<\eta<1.00097
\end{equation}

satisfy the above inequalities (In this case, see Fig. 1b for the resulted ranges of $\gamma$ and $lambda$ for which CCC is violated.). Incorporating all three fundamental constants, ($\hbar$, $c$ and $G$), the inequalities in (\ref{ab}) lead to the following lower and upper bounds for the energy and dimensionless spin parameter of the particle
\begin{equation}
1.28\times10^{-3}\frac{M}{M_{\odot}}<\frac{E}{M_{\odot}c^2}<1.3\times10^{-3}\frac{M}{M_{\odot}}
\end{equation}
\begin{equation}
8.698\times10^{-4}\left(\frac{M}{M_{\odot}}\right)^2<\frac{L}{G{M_{\odot}}^2/c}<8.881\times10^{-4}\left(\frac{M}{M_{\odot}}\right)^2
\end{equation}
where $M_{\odot}$ is the mass of the sun. Likewise,  the quantum number and dimensionless frequency of the field quanta yield
\begin{equation}
7.343\times10^{72}\left(\frac{M}{M_{\odot}}\right)^2<m<7.498\times10^{72}\left(\frac{M}{M_{\odot}}\right)^2
\end{equation}
\begin{equation}
2.36\times10^{73}\left(\frac{M}{M_{\odot}}\right)^2<\frac{\omega}{\Omega}<2.397\times10^{73}\left(\frac{M}{M_{\odot}}\right)^2
\end{equation}
where $\Omega=\sqrt{\gamma}c^2/2Gr_+$ is the angular velocity of the horizon and $m$ is an integer in the above interval. For a typical stellar-mass BH like GRS 1915+105 in the Milky Way, whose mass and dimensionless spin parameter are $M\sim10-18 M_\odot$ and $CJ/GM^2\sim.8-1$, we obtain:
\[
1.28\times10^{-2}<\frac{E}{{M_{\odot}c^2}}<2.34\times10^{-2}
\]
\[
8.698\times10^{-2}<\frac{L}{G{M_{\odot}}^2/c}<2.877\times10^{-1}
\]
\begin{equation}
7.343\times10^{74}<m<2.429\times10^{75}  \hspace{.7cm}, \hspace{.7cm} 2.36\times10^{75}<\frac{\omega}{\Omega}<7.776\times10^{75}
\end{equation}
From above, we see that the particle  would have to have an energy within the  narrow range of 
about $1-2\%$ of the mass of the sun to get 1905+105 to violate CCC.  
This fine tuning also holds for the angular momentum of the particle and also for the parameters of the field.

\begin{figure}
	\begin{center}
		\includegraphics[height=15cm]{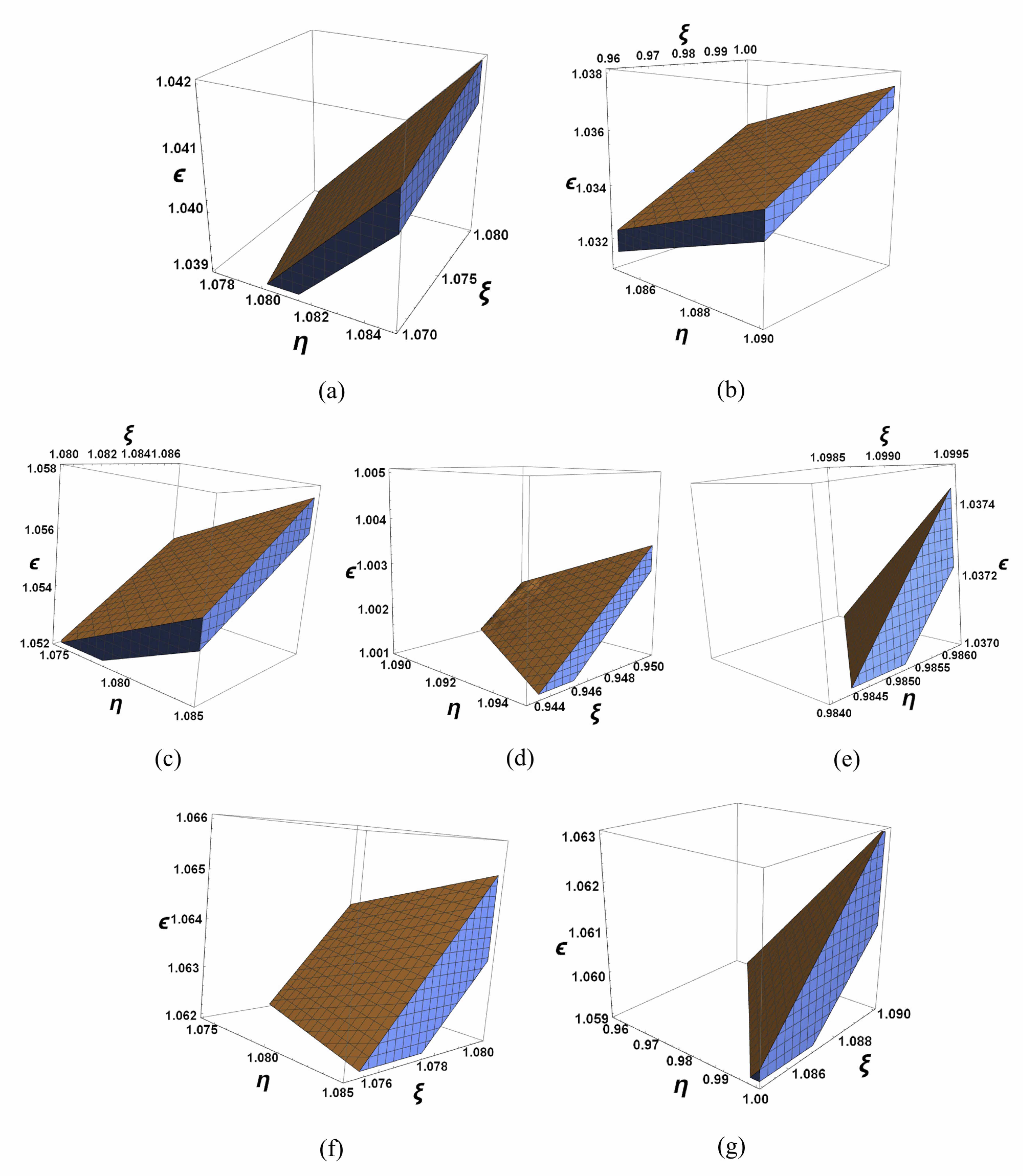}
		\caption{\small{CCC is violated in the specified region of (a)-(g) for a non extremal KN BH and we have set $\gamma+\lambda=0.9898$. In diagrams (a) and (b), $\gamma>\lambda$ and $\gamma= 0.8383$, in diagrams (c), (d) and (e), $\gamma\sim\lambda$ and $\gamma=0.3939$ and in diagrams (f) and (g), $\gamma<\lambda$ and $\gamma=0.2222$.}}
		\label{fig:4}
	\end{center}
\end{figure}

We next consider a non-extremal KN BH for which $\lambda+\gamma<1$ in addition to the inequalities (\ref{4*}) and (\ref{6*}). There are now two input parameters, $\gamma$ and $\lambda$. According to Fig. 1c, we see that the violation of CCC does not occur for $\gamma+\lambda$  sufficiently less than one. As before, the valid ranges of parameters for violation of CCC are thoroughly evaluated and presented in figure (\ref{fig:4}). As with the extremal case, CCC violation again requires fine-tuning for large astrophysical values of the parameters. If the probe and BH rotate in opposite directions and have opposite-signed charges,  CCC remains valid. Moreover, this result still holds for two other  cases. One is when a Kerr-like BH ($\gamma>\lambda$) rotates in the opposite direction to that of the probe but their charges have the same sign. The other  is when an RN-like BH ($\lambda>\gamma$) rotates in the same direction to that of the probe but their charges have opposite signs.

\section{Hoop conjecture}

According to the hoop conjecture \cite{th,yu,YU}, a BH will form when a mass M is compressed into a region whose circumference is smaller than $4\pi M$ in every direction. This means that each probe can contribute to the mass of BH at the hoop radius, $r_{\text{hoop}}=2(M+\delta M)$. For a general KN BH, the hoop radius is generalized in \cite{hod} in which the Schwarzschild event horizon is substituted by the corresponding KN BH one, $r_{\text{hoop}}=M+\sqrt{M^2-a^2-Q^2}$. 
Now suppose that we throw a number of probes{\footnote{A similar argument which allows to discretize the single test body absorption in a series of smaller processes, can be found in \cite{chir}.}, one by one, towards a KN BH.  As mentioned above, according to the hoop conjecture, the event horizon of the KN BH is slightly modified when an appropriate radius is crossed. Thus each particle or quanta  as it falls, sees the KN BH modified  due to its predecessors. 

In the $n$th step, the new event horizon is  
\begin{equation}
r_{{n}_{\text{hoop}}}=M_n+\sqrt{{M_n}^2-{a_n}^2-{Q_n}^2}
\end{equation}
in which $ M_n=M+n\delta M$, $ Q_n=Q+n\delta Q $ and $ a_n=\frac{aM+n\delta J}{M+n\delta M}$ are the parameters of the BH after the $n^{th}$ particle was absorbed by the BH. Following the discussion of section 2, the forms of equations (\ref{4*}), (\ref{6*}) and (\ref{7*}) remain unchanged by defining the following new dimensionless parameters
\begin{equation}
\begin{matrix}
\epsilon_n=1+\frac{\bar{E}}{1+n\bar{E}}&,&\xi_n=1+\frac{\bar{q}}{1+n\bar{q}}&,&\eta_n=1+\frac{\bar{L}}{1+n\bar{L}}\\\\\gamma_n=\frac{(1+n\bar{L})^2}{(1+n\bar{E})^4}\gamma_0&,&\lambda_n=(\frac{1+n\bar{q}}{1+n\bar{E}})^2\lambda_0
\end{matrix}
\label{**}
\end{equation}
where $\bar{E}=E/M$, $\bar{L}=L/aM$ and $\bar{q}=q/Q$. Obviously for $n=0$  we  recover the parameters introduced in section 2. Repeating the discussion of sections 2 and 3, it is straightforward to see that, due to (\ref{**}), the allowed ranges of probe's parameters for violating CCC, becomes progressively narrower as $n$ increases. 

To see this, let us start from the obtained ranges of $ \epsilon_0$, $\eta_0$ and $\xi_0 $ in figure \ref{fig:1}a,  which  corresponds ton an extremal KN BH with $\gamma_0=0.8383$. Substituting these in (\ref{**}), the resulting ranges of $\epsilon_n$, $\eta_n$ and $\xi_n$ decrease as $n$ increases. For example, taking $n=1000$, we get $ 1.000894<\epsilon_{1000}< 1.000898$ , $1.000947 <\eta_{1000}<1.00095 $ and $1.000642 <\xi_{1000}<1.000655 $. These lead to highly fine tuned and very large values of probe parameters in astrophysical terms.  This is also true for other types of BHs, such as Kerr, RN and KN in both extremal and non-extremal cases.

We note that, as expected, in the limit $n\rightarrow\infty$, we have $\epsilon_n\rightarrow 1$, $\eta_n\rightarrow 1$ and $\xi_n\rightarrow 1$. Also in this limit $\lambda_n\rightarrow (\frac{q}{E})^2$ and $M^2\gamma_n\rightarrow (\frac{L}{E})^2$. Remembering the definitions of these parameters, this means that according to the hoop conjecture, if one throws an infinite number of particles or quanta into a KN BH, finally $\frac{Q^2}{M^2}\rightarrow\frac{q^2}{E^2}$ and $\frac{M^2a^2}{M^2}\rightarrow\frac{L^2}{E^2}$. Substituting these values into the relations (\ref{4*}), (\ref{6*}) and (\ref{7*}), one can easily find that the CCC  cannot be violated for a general KN BH when an infinite number of particles or quanta are thrown towards it, which is in confirmation with \cite{13}. 

\section{Concluding Remarks}

 Using a probe that is falling into a general KN BH and neglecting back-reaction, we have shown that the violation of CCC requires large and fine tuned probe parameters. This occurs for an extremal and a near extremal KN BH.  Furthermore, the further from extremality, the more difficult violation of CCC becomes. 
 
 We have classified the allowed ranges of energy, charge and   angular momentum of the probe that yield CCC violation and the results are presented in figures \ref{fig:1} and \ref{fig:4}. 
 We see that the results depend on three factors:  the relative magnitudes of  charge and  angular momentum of the BH ($\gamma$ and $\lambda$), whether the infalling particle or field quanta and BH rotate in the same or opposite directions ($\eta>1$ or $\eta<1$) and whether they have the same or opposite sign of charge ($\xi>1$ or $\xi<1$). Figures \ref{fig:1} and  \ref{fig:4}  are extremely useful in illustrating many interesting points, in particular in which cases the CCC can be violated assuming that the probe has a very small energy (or angular momentum or charge), for example   figure \ref{fig:4}d. 
 
More importantly,  in choosing larger values for the probe parameters, one doesn't obtain a continuous interval  of these parameters for CCC violation and thus more fine tuning is needed. To clarify this point, for an extremal BH we have plotted the inequalities (\ref{12*}) and (\ref{13*}) in the intervals  $ 1.07<\epsilon<1.08 $, $ 1.09<\xi<1.1 $ and $ 0.02<\eta <1.1 $  for $ \gamma=0.2222$ in figure \ref{fig:2}a. A similar situation holds for moving away from near extremal BH as $\gamma+\lambda$ decreases, illustrated in figures  \ref{fig:2}b and \ref{fig:2}.c.  For this case, the parameters should be larger and more fine tuning is needed.  
In astrophysical terms  the allowed values of  the probe  parameters, measured in solar mass units, is indeed large.  When the black hole mass is tiny the values likewise become tiny but still fine-tuned.  

\begin{center}
	\begin{figure}
		\includegraphics[height=12cm]{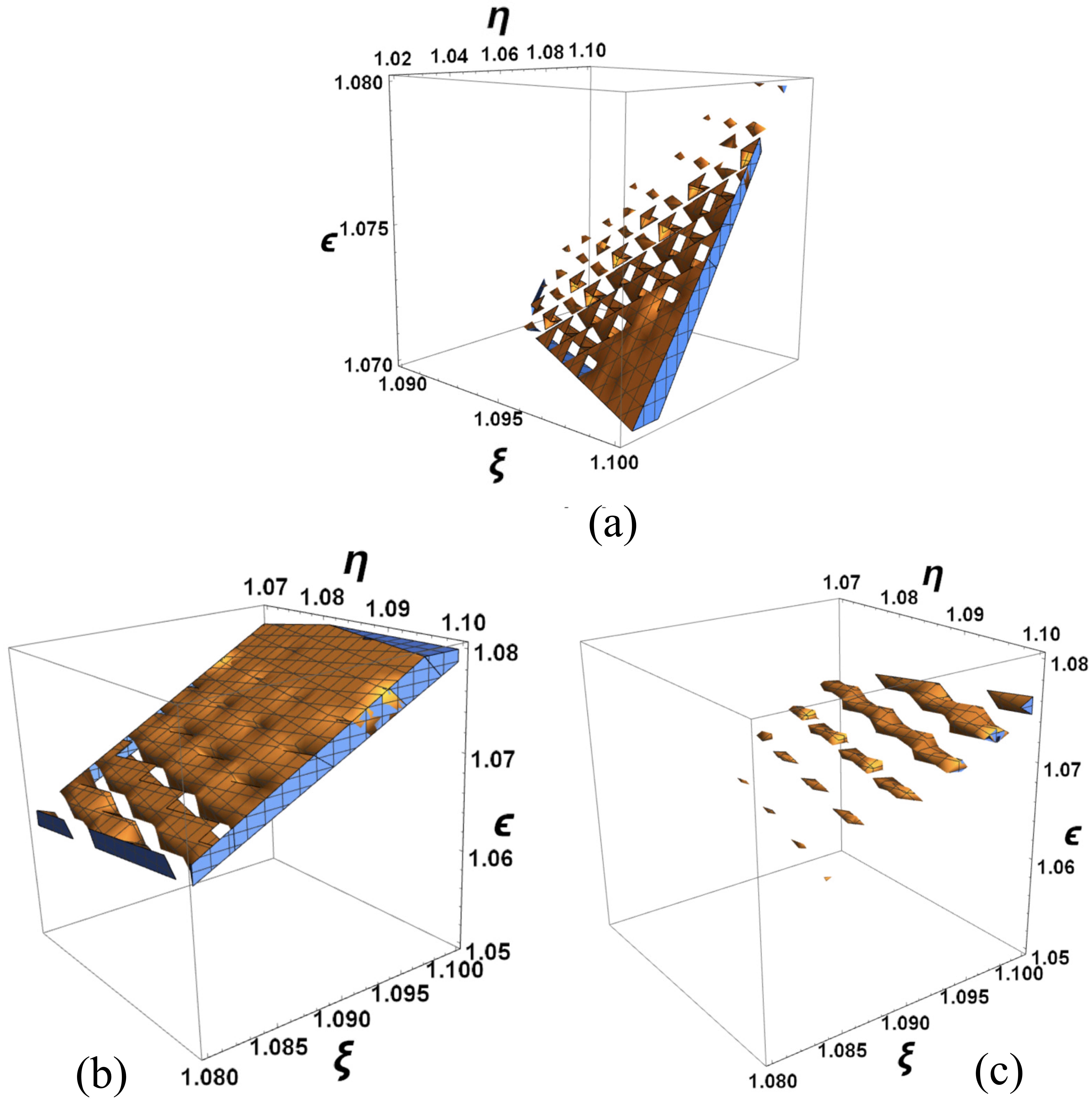}
		\caption{(a): \small{Inequalities (\ref{12*}) and (\ref{13*}) for $ \gamma\simeq 0.2222$ in intervals $ 1.07<\epsilon<1.08 $, $ 1.09<\xi<1.1 $ and $ 1.02<\eta<1.1 $}
			\small{(b) and (c): Inequalities (\ref{4*}) and (\ref{6*}) in the intervals $ 1.05<\epsilon<1.08$, $ 1.08<\xi<1.1 $ and $ 1.07<\eta<1.1 $, (b): $ \gamma\simeq 0.1515$ and $ \lambda\simeq 0.8383$ $ (\gamma+\lambda=0.9898)$ and (c):  $ \gamma\simeq 0.1616$ and $ \lambda\simeq 0.8181$ ($ \gamma+\lambda=0.9797 $)}
		}
		\label{fig:2}
	\end{figure}
\end{center}

We have employed the  hoop conjecture to show that the greater the number of infalling particles or field quanta, the narrower the range of allowed parameters for CCC violation. As a result, as the number of particles or quanta approaches infinity, CCC violation is not possible for both extremal and non extremal Kerr, RN and KN BHs.

\section*{Acknowledgements}
The authors would like to thank the Iran National Science Foundation (INSF) for supporting this research under grant number 97015575. F. Shojai is grateful to the University of Tehran for supporting this work under a grant provided by the university research council.

\end{document}